\documentclass[aps,twocolumn,floats,prd,nofootinbib,superscriptaddress,10pt]{revtex4-1}

\usepackage[utf8]{inputenc}
\usepackage[T1]{fontenc}
\usepackage{natbib}
\usepackage{amsmath}

\usepackage{graphicx}
\usepackage{dcolumn}
\usepackage{bm}
\usepackage{hyperref}

\makeatletter
\newcommand{\PBH}{\text{PBH}}
\renewcommand{\i}{\text{i}}
\makeatother

\begin{document}


\title{Constraints on the primordial curvature power spectrum from primordial black holes}

\author{Gabriela Sato-Polito}
\affiliation{%
 Department of Physics \& Astronomy, Johns Hopkins University,\\
 3400 N. Charles St., Baltimore, MD 21218, USA
}%

\author{Ely D. Kovetz}%
\affiliation{%
 Department of Physics \& Astronomy, Johns Hopkins University,\\
 3400 N. Charles St., Baltimore, MD 21218, USA
}%
\affiliation{Department of Physics, Ben-Gurion University of the Negev, Be'er Sheva 84105, Israel}

\author{Marc Kamionkowski}
\affiliation{%
 Department of Physics \& Astronomy, Johns Hopkins University,\\
 3400 N. Charles St., Baltimore, MD 21218, USA
}%


\date{\today}

\begin{abstract}
Large-amplitude density perturbations may have collapsed during the radiation dominated epoch of the Universe to form primordial black holes (PBHs). There are several constraints to the abundance of PBHs that stem from evaporation or gravitational effects. Due to the connection between primordial perturbations and the formation of PBHs, constraints on the present-day abundance of PBHs can be translated into limits on the primordial curvature power spectrum. We introduce several new observational and forecasted constraints to the amplitude of the primordial power spectrum and incorporate in our analysis uncertainties in the critical overdensity for collapse and considerations of ellipsoidal collapse. Our results provide the most stringent limits from PBHs on the primordial curvature power spectrum on small scales.
\end{abstract}

\maketitle


\section{\label{sec:level1} Introduction}
Primordial black holes (PBH) may have formed in the early Universe through the collapse of density perturbations that were produced during inflation \cite{zeldovich1967, hawking1971, carr1975, chapline1975}. The amplitude of these scalar perturbations has been measured through observations of the cosmic microwave background (CMB) on large scales ($10^{-4}\ $Mpc$^{-1} \lesssim k \lesssim 1 \ $Mpc$^{-1}$) to be of order $P_{\mathcal{R}} \sim 2 \times 10^{-9}$ \cite{planck2018}. On scales smaller than those probed by the CMB, limits on the abundance of PBHs can offer useful constraints on the primordial curvature power spectrum and therefore on models of inflation.

After the end of inflation, the Hubble horizon begins to grow and density perturbations re-enter the horizon. The probability that a region will collapse to form a PBH is determined by its density contrast at horizon re-entry. In order to form PBHs, the critical overdensity $\delta_c$ for collapse requires the primordial curvature power spectrum to be around seven orders of magnitude higher on small scales than on large scales. A variety of inflationary models have been studied where an increase in power on small scales can occur \cite{ivanov1994, carr1994, leach2000, garcia1996, yokoyama1995, yokoyama1998, kawasaki1999, leach2000, chongchitnan2007, kawaguchi2008, saito2008, frampton2010, kawasaki2013, kohri2013, clesse2015, kawasaki2016, murgia2019}.


The connection between PBH formation and primordial perturbations allows us to convert between constraints on the abundance of PBHs and constraints on the primordial curvature power spectrum \cite{josan2009, bugaev2008, emami2018}. There are several limits on the abundance of PBHs that stem from evaporation or from present day gravitational effects. PBHs with masses $ \lesssim 10^{15} $g will have evaporated by the present day due to Hawking radiation and their abundance can be constrained by the effects of the emitted particles on the gamma-ray background~\cite{macgibbon1991} and on big-bang nucleosynthesis (BBN)~\cite{kohri1999}. The abundance of heavier PBHs that survive to the present day can be constrained by a variety of effects, such as gravitational lensing, dynamical effects, and, more recently, through gravitational waves~\cite{Kovetz:2017rvv, carr2010,  graham2015, tisserand2007, niikura2017, mediavilla2017, niikura2019, zumalacarregui2018, lorimer2007, munoz2016, ji2018, venumadhav2017, schutz2017, ali2017, raidal2017, brandt2016, haimoud2017, clark2017, gaggero2016, griest2013, katz2018, capela2013} (for a perspective on the future of the field, see Ref. \cite{haimoud2019}).

In this paper we revisit the current observational limits on the abundance of PBHs and provide state-of-the-art constraints on the primordial curvature power spectrum. We introduce several novel current and forecasted constraints from the strong lensing of fast radio bursts~\cite{munoz2016}, highly magnified caustic-crossing stars~\cite{venumadhav2017}, microlensing of stars in the Galactic bulge~\cite{niikura2019}, Shapiro time delays in pulsar timing experiments for known pulsars and those expected to be detected by SKA, the binary black hole merger rates set by LIGO O1~\cite{ali2017}, and the non-detection of a stochastic gravitational wave background~\cite{raidal2017}. In addition, we address the uncertainties in the determination of $\delta_c$, the critical overdensity threshold for collapse, which was shown to depend on the density perturbation profile, as well as the effect of non-spherical collapse~\cite{harada2013, musco2005, akrami2016, kuhnel2016}.

This paper is organized as follows: in Section \ref{sec:abundance} we outline the PBH formation mechanism assumed throughout this paper and relate the mass fraction of PBHs at the formation time to the present-day energy density of PBHs. In Section \ref{sec:constraints} we briefly discuss the observational and forecasted constraints on the fraction of dark matter in the form of PBHs included in our analysis. In Section \ref{sec:ps} we detail the conversion of PBH abundance constraints to limits on the primordial curvature power spectrum, and discuss a few uncertainties on this conversion in Section \ref{sec:uncertainty}. We present our main results in Section \ref{sec:results} and conclude in Section \ref{sec:conclusions}.

\section{Primordial black hole abundance \label{sec:abundance}}
We assume that PBHs form directly from density perturbations that collapse during the radiation-dominated epoch and have a monochromatic mass function (see, e.g., Ref \cite{carr2017} for a comprehensive study of extended mass functions), with masses equal to a fixed fraction $\gamma$ of the horizon mass, which is determined by the details of the gravitational collapse. That is,

\begin{equation}
  M_{\PBH}(k) = \gamma \left. \frac{4 \pi}{3} \rho H^{-3} \right|_{k=aH},
\end{equation}
where $M_{\PBH}$ is the mass of the PBH that is formed when a scale $k$ reenters the horizon and $\gamma \equiv 0.2$ \cite{carr1975}.

The mass of the PBHs can be related to the present-day horizon mass $M_0 = 4 \pi \rho_{\text{cr}}/(3 H^3_0) \simeq 4.7 \times 10^{22} M_{\odot}$ by using the assumption that they form during a radiation-dominated epoch and the conservation of entropy. Since the entropy density $s$ is given by $s = g_{*s} a^3 T^3$ and the radiation density is $\rho = \frac{\pi^2}{30} g_* T^4$, an expansion at constant entropy implies\footnote{We have taken the number of entropy degrees of freedom $g_{*s}$ and the total number of relativistic degrees of freedom $g_*$ to be approximately equal.} $\rho \propto g^{-1/3}_* a^{-4}$ . Labeling the quantities evaluated at the time of PBH formation by "$\i$" and the present-day value by "$\text{0}$", we have that

\begin{equation}
  \rho^{\i}_{\text{tot}} = \left( \frac{g^{\text{0}}_*}{g^{\i}_*} \right)^{1/3} a^{-4}_{\i} \rho_{\text{cr}} \Omega_r,
\end{equation}
where the present-day radiation density $\Omega_r$ is approximately $9\times 10^{-5}$, and the present-day critical density $\rho_{\text{cr}}$ is given in terms of the Hubble constant $H_0$ as $\rho_{\text{cr}} = 3 H^2_0/8\pi G$.

The mass of the PBH can then be calculated as

\begin{equation}
  M_{\PBH} = \gamma \Omega_r^{1/2} M_0 \left( \frac{g^0_*}{g^{\i}_*}\right)^{1/6} \left(\frac{H_0}{k}\right)^2 .
\end{equation}

The initial mass fraction $\beta$ of primordial black holes is related to the present day PBH density $\Omega_{\PBH}$ as

\begin{equation}
  \beta \equiv \frac{\rho_{\PBH}^{\i}}{\rho_{\text{tot}}^{\i}} = \left(\frac{\Omega_{\PBH}}{\Omega_r}\right) \left(\frac{g^{\i}_*}{g^0_*}\right)^{1/3} a_{\i}.
\end{equation}
The scaling as $a_{\i}$ arises from the fact that the PBH density scales as $a^3$ and $\rho_{\text{tot}}^{\i}$ as $a^4$.

The initial mass fraction can be rewritten using the results above as (see, e.g., Ref. \cite{nakama2017} for a complete derivation)

\begin{equation}
\begin{split}
\beta &= \left(\frac{\Omega_{\PBH}}{\Omega_r^{\frac{3}{4}}}\right) \left( \frac{M_{\PBH}}{M_0}\right)^{1/2} \gamma^{-\frac{1}{2}} \left(\frac{g^{\i}_*}{g^0_*}\right)^{\frac{1}{4}} \\ &\simeq 4 \times 10^{-9} \left( \frac{\gamma}{0.2} \right)^{-\frac{1}{2}} \left(\frac{g^{\i}_*}{10.75}\right)^{\frac{1}{4}} \left( \frac{M_{\PBH}}{M_{\odot}} \right)^{\frac{1}{2}}  f_{\PBH},
\end{split}
\end{equation}
where we define the fraction $f_{\PBH}$ as the fraction of dark matter that is made up of PBHs

\begin{equation}
  f_{\PBH} \equiv \frac{\Omega_{\PBH}}{\Omega_{\text{DM}}},
\end{equation}
and $\Omega_{\text{DM}}$ is the present-day dark matter density.

\section{Constraints on the PBH fraction \label{sec:constraints}}

{\bf Cold dark matter density.} The simplest constraint on the present day PBH density is that it must be smaller than the cold dark matter density, which corresponds to

\begin{equation}
  f_{\PBH} < 1.
\end{equation}

This only applies to PBHs with masses above $\simeq 10^{15} g$ that have not evaporated by the present day.

{\bf Extragalactic gamma ray background.} PBHs with $10^{15} g \lesssim M_{\PBH} \lesssim 10^{17} g$ remain to the present day and would be emitting a significant flux of gamma rays. Their abundance can be constrained by the contribution of gamma-ray emission to the extragalactic gamma-ray background, using observations from EGRET and Fermi LAT \cite{carr2010}.

{\bf White dwarfs.} As pointed out by Ref. \cite{graham2015}, PBHs could also disrupt a white dwarf as they pass through it, which would lead to an ignition of a thermonuclear runaway and trigger a supernova explosion. The observed distribution of white dwarfs was used to place constraints on PBHs with $10^{19} g \lesssim M_{\PBH} \lesssim 10^{20} g$.

\begin{figure}[h]
  \centering
    \includegraphics[width=0.5\textwidth]{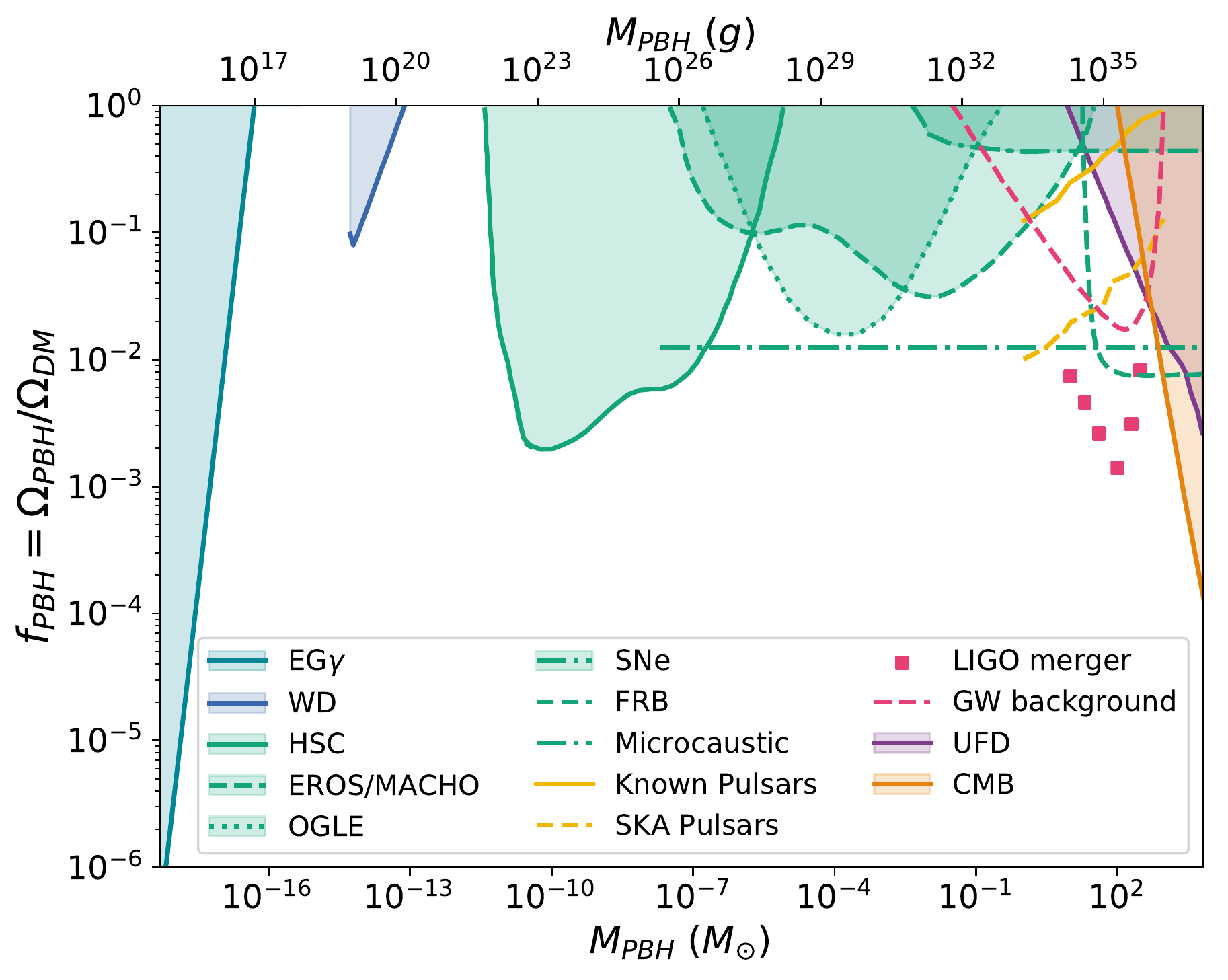}
    \caption{Observational (shaded) and forecasted (un-shaded) constraints to the fraction of PBHs to dark matter. These include observations of the extra-galactic gamma ray background (EG$\gamma$), constraints from white dwarves (WD), lensing events (HSC, EROS/MACHO, SNe, OGLE), ultra-faint dwarf galaxies (UFD), the cosmic microwave background (CMB), and forecasts for observation of lensing of fast radio bursts (FRB), of caustic-crossing stars (Microcaustic), pulsar timing (known pulsars and SKA pulsars), the LIGO merger rate, and the stochastic gravitational wave background. We leave out additional constraints that are weaker than the observational limits presented here.}
    \label{fig:f}
\end{figure}

{\bf Gravitational lensing.} Microlensing events can be used to place constraints on the abundance of PBHs with masses above $10^{22} g$. The EROS and MACHO surveys observed stars in the Large and Small Magellanic Clouds and the null detection of microlensing was used to constrain PBHs with $10^{26} g \lesssim M_{\PBH} \lesssim 10^{34} g$ \cite{alcock2000, tisserand2007}. The non-detection of PBH-induced microlensing events of stars in the Andromeda galaxy (M31), observed with the Subaru Hyper Suprime-Cam (HSC), placed stringent upper bounds on the abundance of PBHs with $10^{22} g \lesssim M_{\PBH} \lesssim 10^{28} g$ \cite{niikura2017}. More recently, Ref. \cite{niikura2019} constrained the abundance of PBHs with masses $10^{27} g \lesssim M_{\PBH} \lesssim 10^{30} g$ through microlensing searches in 5 years of observations of stars in the Galactic bulge with the Optical Gravitational Lensing Experiment (OGLE). The abundance of PBHs was also constrained by the absence of an effect on the lensing PDF of type Ia supernovae for masses $\gtrsim 10^{31}g$ \cite{zumalacarregui2018} (see Ref. \cite{garcia2017} for a counterpoint). Observations of gravitationally lensed quasars can also be used to place constraints on the fraction of PBHs to dark matter \cite{mediavilla2017}. However, we do not include this limit in Figure \ref{fig:f}, since it is less constraining than other measurements in the same mass range.

{\bf Lensing of FRBs and GRBs.} Fast radio bursts (FRBs) are radio pulses with millisecond durations \cite{lorimer2007}. Ref. \cite{munoz2016} proposed to use the strong lensing of FRBs to probe PBHs with masses $ \gtrsim 10 M_{\odot}$. FRBs that are strongly lensed by PBHs would produce two images of the burst. Although the angular separation between the images might not be resolved, the time delay between them can be, since lenses with masses above $\sim 10 M_{\odot}$ will generate a time delay larger than the pulse width. The dashed green line in Fig. \ref{fig:f} shows the forecast of one year of observations for an experiment like CHIME \cite{chime2014}. The constraint shown corresponds to the smallest fraction $f_{\PBH}$ that would produce at least one lensing event with a time delay longer than 1 ms, the typical intrinsic width of an FRB~\cite{Petroff:2016tcr}. Ref.~\cite{ji2018} investigated using gamma-ray-burst (GRB) light curves to search for echoes induced by compact dark matter, but concluded that current experiments cannot achieve the necessary signal-to-noise for the echo to be detectable.

{\bf Caustic perturbations.} Distant galaxies can be lensed by galaxy clusters. In a smooth lens model, clusters have critical curves where the magnification is formally infinite. As a star that belongs to the lensed galaxy moves outward and crosses the caustic curve, its two images move closer together and merge. The presence of compact objects such as intracluster stars or compact dark matter can disturb the smooth caustic and produce a network of corrugated microcaustics. Ref. \cite{venumadhav2017} studied the lensing properties of a cluster with a granular mass density and proposed using the peak magnification and frequency of caustic-crossing events to probe dark matter. In the vicinity of the smooth critical curve produced by the cluster, the microcritical curves of the point masses will merge and form a band of width $2r_w$. The constraint shown in Fig. \ref{fig:f} corresponds to the smallest fraction of PBHs such that they dominate the width of the microcaustic network $r_w$ over the intracuster stars. For an abundance greater than the green dash-dotted line, $r_w$ is larger than what would be expected from intracluster stars. We emphasize, however, that this forecast is for idealized observations, rather than any specific observational program.

{\bf Pulsar timing array experiments.} The passage of a PBH near the line of sight between the Earth and a pulsar would shift the pulse arrival times. Pulsar timing can be used to constrain the PBH abundance via the non-detection of third-order Shapiro time delay in the mass range $\sim 1-10^3 M_{\odot}$ \cite{schutz2017}. Fig. \ref{fig:f} includes the projected constraints from known pulsars and from future SKA pulsars.

{\bf LIGO merger rates.} If PBHs can form binaries in the early Universe~\cite{Sasaki:2016jop} that remain bound to the present day, their merger rate would dominate that of PBH binaries formed in the late Universe~\cite{Bird:2016dcv}. Ref.~\cite{ali2017} constrained the PBH abundance in the mass range $\sim 10-300 M_{\odot}$  by requiring that the merger rate of PBHs does not exceed the upper limit on the merger rate from the O1 LIGO observing run~\cite{TheLIGOScientific:2016pea}. Meanwhile, Ref.~\cite{raidal2017} derived constraints on the PBH abundance from the non-observation of a stochastic gravitational wave background.

{\bf Dynamical effects.} The presence of massive compact objects can disrupt astrophysical systems through gravitational interactions. If PBHs are present in ultra-faint dwarf (UFD) galaxies, they would dynamically heat the stellar population. Ref. \cite{brandt2016} used the stellar distribution of UFD galaxies to constrain PBHs with $10^{34} g \lesssim M_{\PBH} \lesssim 10^{37} g$.

{\bf Cosmic Microwave background.} PBHs from the primordial plasma could accrete during recombination. The produced radiation would then affect the CMB temperature and polarization spectrum. The most conservative constraints from Planck 2015 \cite{planck2015} CMB data were presented in Ref. \cite{haimoud2017} (see also Refs.~\cite{ricotti2008, Blum:2016cjs, poulin2017}, and Ref.~\cite{murgia2019} for constraints from the Lyman-alpha forest).

\section{Constraints on the primordial curvature power spectrum \label{sec:ps}}
A region will collapse to form a PBH if its density contrast at horizon re-entry is above a certain threshold $\delta_c$. Assuming Gaussian initial perturbations, the probability density of the smoothed density contrast $\delta(R)$ is given by

\begin{equation}
  P(\delta(R)) = \frac{1}{\sqrt{2\pi} \sigma(R)} \exp \left( - \frac{\delta^2(R)}{2\sigma^2(R)}\right),
\end{equation}
where $R = (aH)^{-1}$, and the mass variance is given by

\begin{equation}
  \sigma^2(R) = \int_0^{\infty} \tilde{W}^2(kR) P_{\delta}(k) \frac{dk}{k},
\end{equation}
where $P_{\delta}$ is the power spectrum of the matter density field and $\tilde{W}$ is the Fourier transform of a Gaussian window function,
$ \tilde{W}(kR) = \exp \left( - k^2 R^2/2 \right)$.

In Press-Schechter theory, the relative abundance of primordial black holes of mass $M_{\PBH}$ is equivalent to the probability that the smoothed density field exceeds the threshold $\delta_c = 0.42$ \cite{harada2013},

\begin{equation}
  \beta(M_{\PBH}) = 2\int^1_{\delta_c} d\delta(R) P(\delta(R))
  \approx \text{erfc} \left( \frac{\delta_c}{\sqrt{2} \sigma(R)} \right).
\end{equation}
This expression can be inverted to translate the constraint on the initial mass fraction of PBHs into a constraint on the mass variance and therefore the density power spectrum $P_{\delta}$. Converting the density to the curvature perturbation $\mathcal{R}$, we get

\begin{equation}
  \delta(k, t) = \frac{2(1 + \omega)}{5 + 3\omega} \left( \frac{k}{aH} \right)^2 \mathcal{R},
\end{equation}
with $\omega = 1/3$ for radiation domination. The primordial curvature power spectrum is therefore

\begin{equation}
  P_{\delta} (k, t) = \frac{4(1 + \omega)^2}{(5 + 3\omega)^2} \left( \frac{k}{aH} \right)^4 P_{\mathcal{R}}(k).
\end{equation}

We constrain the primordial power spectrum by assuming that it is scale invariant at each $k$, since the integral in $\sigma^2$ is dominated by the scale $k = R^{-1}$. That is, for each scale $k$ constrained by the abundance of PBHs with mass $M_{\PBH}$, we have that $P_{\mathcal{R}} \propto \sigma^2(R)$.

\section{Uncertainties on the primordial power spectrum constraint \label{sec:uncertainty}}
The method described thus far neglects well-known uncertainties in the conversion between constraints on the fraction of PBHs and the primordial curvature perturbations \cite{akrami2016, carr2016, sasaki2018}. Here we discuss how our constraints are weakened by the uncertainty in determining $\delta_c$ and the effect of non-spherical collapse.

The collapse of PBHs has been extensively studied analytically and numerically in order to determine the value of the density perturbation threshold for PBH formation \cite{niemeyer1999, shibata1999, green2004, musco2005, polnarev2007, harada2013}. Numerical simulations have shown, however, that the precise value of $\delta_c$ depends on the density or curvature perturbation profile. Perturbations with different shapes lead to different density thresholds for collapse that vary between 0.42 and 0.66 \cite{polnarev2007}.

Departures from spherical symmetry are expected in a more realistic shape distribution of the primordial perturbations. Ref. \cite{kuhnel2016} investigated the effect of ellipsoidal collapse on PBH production and found a significant decrease, leading to weaker constraints on the primordial curvature power spectrum. This can be shown by using the ellipsoidal collapse threshold \cite{sheth2001} obtained in the context of the collapse of dark matter halos. The threshold  $\delta_{ec}$ for an ellipsoidal density perturbation to collapse is given in terms of the value for spherical collapse as

\begin{equation}
  \delta_{ec} = \delta_c \left[ 1 + \kappa \left( \frac{\sigma^2}{\delta^2_c} \right)^{\nu} \right],
\end{equation}
with $\kappa = 9/\sqrt{10 \pi}$ and $\nu = 1/2$.

\begin{figure}
  \centering
    \includegraphics[width=0.5\textwidth]{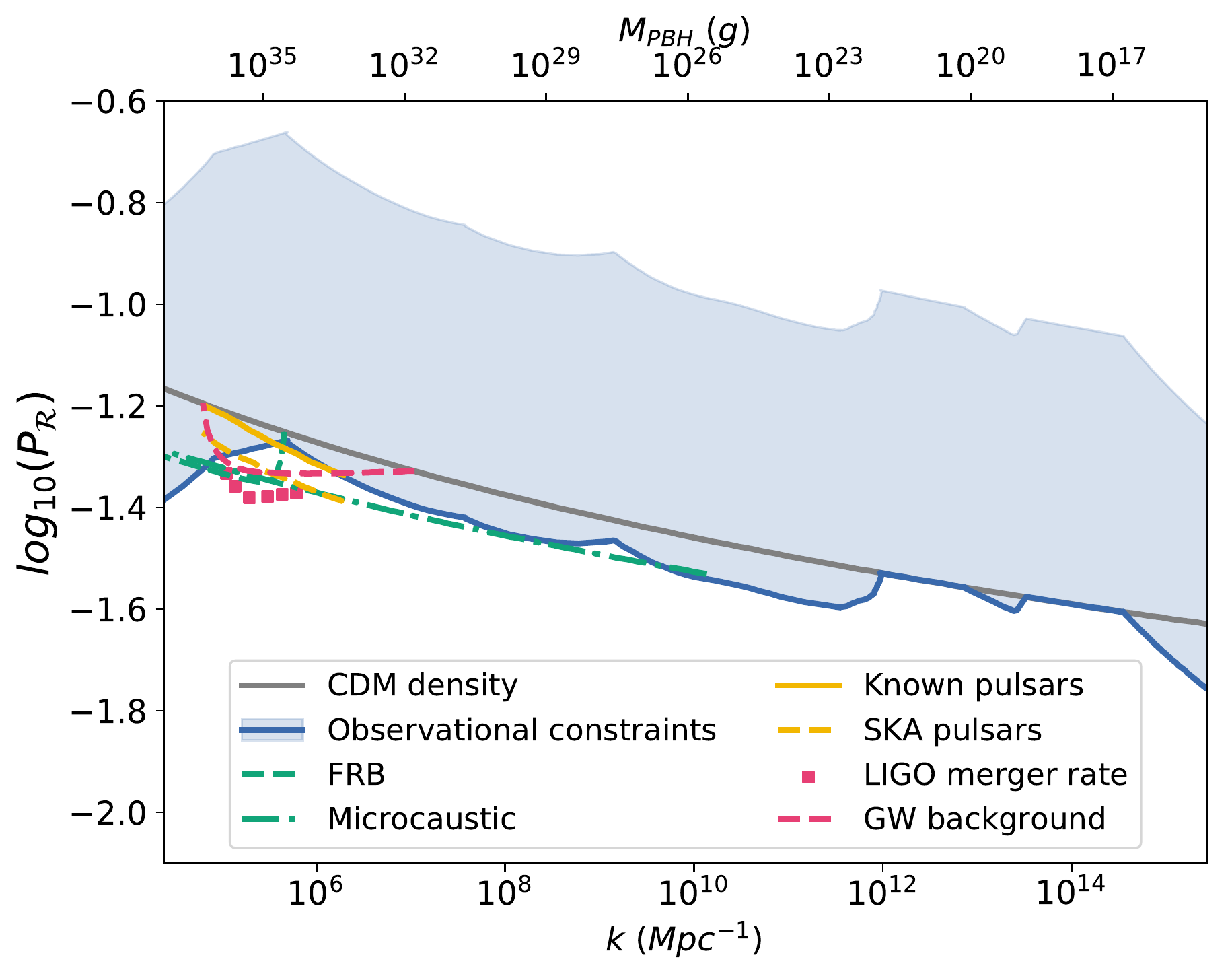}
    \caption{Constraints on the primordial curvature power spectrum by the abundance of PBHs. We show the most promising forecasted constraints as distinctly labeled curves, which correspond to FRB lensing (FRB), caustic-crossing stars (Microcaustic), pulsar timing (known pulsars and SKA pulsars), the merger rate from LIGO, and the stochastic gravitational wave background. The current observational constraints were merged and the most stringent bound is shown in solid blue. The range of uncertainty regarding the current constraints is shown in shaded blue, and the limit that the density of PBHs cannot exceed that of dark matter is the solid grey line.}
    \label{fig:P}
\end{figure}

\section{Results \label{sec:results}}
Our results are summarized in Fig. \ref{fig:P}. All limits shown in Fig. \ref{fig:f} were converted into constraints on the primordial curvature power spectrum. We highlight the most promising forecasted constraints by showing them as distinct curves in Fig. \ref{fig:P}. The solid blue line corresponds to the observational constraints, where spherical collapse was assumed with $\delta_c = 0.42$. The blue shaded region is the range of possible values for such constraints, given the uncertainties in $\delta_c$ described above.

One can also consider the constraint that could be derived if we could somehow infer that no PBH exists in the entire observable Universe. A detailed description is given in Ref. \cite{cole2018}, where the limit is obtained by assuming that no regions in the early Universe were dense enough to form a PBH. This calculation yields a constraint of order $\sim 10^{-2}$, but we omit this line in Fig \ref{fig:P}.

We note that by including the uncertainty in $\delta_c$, the forecasted constraints shown in Figure \ref{fig:P} would be shifted as well. We only show the uncertainty in the current observational constraints for the sake of clarity.

\section{Conclusions \label{sec:conclusions}}
In this paper we studied constraints on the primordial curvature power spectrum from non-detection of primordial black holes. The most up-to-date limits on the fraction of dark matter in PBHs were compiled in  Fig.~\ref{fig:f}. Several new bounds were incorporated here for the first time, based on: future observations of FRB lensing in a CHIME-like experiment;  highly magnified stars crossing a network of corrugated microcaustics produced by PBHs; microlensing events from observations of stars in the Galactic bulge by OGLE; forecasts for the non-detection of third-order Shapiro time delay for known and SKA pulsars; the merger rate from LIGO O1; and the stochastic gravitational wave background.

The constraints on the fraction of PBHs as dark matter were then translated into constraints on the primordial curvature power spectrum. We included in our analysis the uncertainty in the determination of the density threshold for collapse and the effect of ellipsoidal collapse, which could significantly weaken the constraint, as indicated in Fig.~\ref{fig:P}. Our results provides the most stringent bounds on scales corresponding to PBHs with masses between $10^{-19}-10^{4}$M$_{\odot}$.

\acknowledgements
We thank Pippa Cole and Tommi Tenkanen for very useful discussions. This work was supported at Johns Hopkins by NASA Grant No.~NNX17AK38G, NSF Grant No.~1818899, and the Simons Foundation.

\bibliographystyle{h-physrev}
\bibliography{references}

\end{document}